# Model of cell response to α-particle radiation

L. J. Liu

Department of Physics, University of Windsor, Windsor, Ont., Canada N9B 3P4

Abstract

Starting from a general equation for organism (or cell system) growth and attributing additional cell death rate (besides the natural rate) to therapy, we derive an equation for cell response to α radiation. Different from previous models that are based on statistical theory, the present model connects the consequence of radiation with the growth process of a biosystem and each variable or parameter has meaning regarding the cell evolving process. We apply this equation to model the dose response for α-particle radiation. It interprets the results of both high and low linear energy transfer (LET) radiations. When LET is high, the additional death rate is a constant, which implies that the localized cells are damaged immediately and the additional death rate is proportional to the number of cells present. While at low LET, the additional death rate includes a constant term and a linear term of radiation dose, implying that the damage to some cell nuclei has a time accumulating effect. This model indicates that the oxygen-enhancement ratio (OER) decreases while LET increases consistently.

## 1. Introduction

When ionizing radiations such as x-rays or α-particles interact with cells in a bio-system, some kinetic energy is being deposited in the system causing possible damage. Linear energy transfer (LET) is the measure of the energy deposited per unit length (kev/μm) of the radiation/particles' track [1]. Depending on the value of LET, ionizing radiation may be divided into two types: low and high LET emissions. For cell response to a low LET radiation such as x-ray, the linear quadratic (LQ) model is successfully used [1,2]. Many survival curves fit the LQ or two component LQ model [2].

Alpha particle emission differs from low LET ionizing radiations in that it belongs to high LET radiation and causes distinct cell response, which has a linear survival curve in most cases [1,3,4]. Barendsen et al. [3] studied human cell response and corresponding oxygen-enhancement ratios (OER, which is the ratio of doses administered under hypoxic to aerated conditions needed to achieve the same biological effect) to α-particles with different energy. Instead of using air, Barendsen et al. used nitrogen for hypoxic condition. The results are summarized in table 1.

**Table 1 Biological effect of human cells after α-particles radiation**

| Energy (Mev) | 2.5 | 3.4 | 4.0 | 5.1 | 8.3 | 25 |
|---|---|---|---|---|---|---|
| LET (kev/μm) | 166±20 | 140±20 | 110±10 | 88±6 | 61±5 | 26±2 |
| OER | 1.0±0.1 | 1.1±0.05 | 1.3±0.1 | 1.7±0.15 | 2.05±0.25 | 2.4±0.3 |
| $\log_{10}(SF) \sim D$ fit | linear | linear | linear | linear | linear | LQ |

LET increases to 166kev/μm while the kinetic energy of each particle decreases to 2.5Mev. Correspondingly, OER decreases to 1.0. The survival curves fit almost linearly when the kinetic energy is less than 8.3 Mev and LET is greater than 61kev/μm. An α-particle has a typical energy of 5 Mev (usually between 3-7 Mev) and is composed of two protons and two neutrons. Relative to other particles such as neutrons, protons and electrons, α-particle is of a large size and mass, which make it have low penetration depth. Therefore, the energy of α-particles is absorbed by tissues (cells) within a short range. This makes it extremely dangerous once the source is ingested or inhaled since all the particles stay. The damage of biological effects is about 20 times as that caused by an equivalent amount of gamma or beta radiation (e.g. number of cells killed in 1 Gy dose). While interacting with cells, α-particle transfers most of its energy in a small region. Therefore, the localized DNA damage is difficult to repair or is even irreparable. Studies [1, 3] show that the OER is 1.0 when LET is greater than 165 kev/μm. Generally, the OER decreases with the increase of LET [5]. When LET is between 61 and 110 kev/μm, the OER is from 2.0 to 1.3, though the survival response is still linear. When LET reduces to 26 kev/μm, the OER increases to 2.4 and the survival curve shows features of the LQ model.

Assuming that a double strand break in the DNA helix is the critical damage, Chadwick and Leenhouts statistically derived the LQ equation and approximated it as [16]:

$$SF = \exp[-k_1 \Delta D - k_2 (1-\Delta)^2 D^2]$$

where $SF$ represents cell surviving fraction, $k_1$, $k_2$ are two constants, $D$ is the absorbed dose; $\Delta$ is a proportion of dose $D$ that is inactivated via single event killing (meaning both strands of the DNA double helix are broken in one radiation event), and $1-\Delta$ is the proportion that is inactivated via double event killing (meaning each strand of DNA double helix is broken independently during different radiation events). Here $\Delta$ is LET related. This model implied that $\Delta$ increases when LET increases. The linear term $\exp(-k_1\Delta D)$ dominates survival at low doses. With increase in LET or $\Delta$, the quadratic term $\exp[-k_2(1-\Delta)^2 D^2]$ plays an increasing role [16]. The LQ model may be applied for cell response to high LET radiations such as α-particles if the quadratic term is ignored.

However, for cell response to α radiation, there are some experimental data that do not fit the LQ model (including the linear limitation) and its multi-component forms. Many survival curves fit a linear line and some might follow the LQ model, while other survival curves do not belong to either of these two cases. For example, Hieber et al. [6] showed that the response of C3H 10T1/2 cells to α radiation deviates upward from the linear line after 1.5 Gy. Beaton et al.'s experiment for A-549s's cell response to α radiation [7] showed a similar result. According to the description for setting up the experiments, the results could not be merely explained as "unattached mitotic cells not reached by the α-particles". From the experimental result given by Durante [8] for H184B5 F5-1 M/10 cell response to α-particles, we can also summarize that the curve is different from either the linear or the LQ model. Some additional experiments [4, 9-13] also showed that the survival curves deviate upwards from a linear line. Some experimental results, such as the survival of asynchronous V79 cells vs. DNA-incorporated activity [14] after 30-min exposure to $^{211}$AtdU (5-[211At]astato-2'-deoxyuridine), may be explained by introducing the two-component exponential model [14, 15]. However, the experimental data in references [6] and [7] do not fit the linear model, its two-component form or any other previous model. Explaining all these survival curves consistently may lead us to propose a general model for cell response to radiation. In fact, all previous models are derived from statistical theory. In the LQ model, the surviving fraction is expressed as $SF = e^{-\alpha D - \beta D^2}$, where $\alpha$ and $\beta$ are two parameters that are determined by cell response. It is assumed that the factor $e^{-\alpha D}$ comes from single event killing, and the factor $e^{-\beta D^2}$ is derived from double event killing [1, 16]. This implies that high LET radiations such as α-particles cause mainly single event killing. In fact, alpha radiation causes more serious damage to cell nuclei (mainly DNA) than other kind of radiations. High LET radiation increases the complexity of lesions due to the formation of multiply damaged sites [17]. The damage includes DNA double strand breaks, which are considered important with evidence relating to cell lethality [17, 18]. In addition, when dose or dose rate is zero, a model for cell response to therapy should be compatible with the natural growth or death process before treatment is applied. However, previous models did not relate to this aspect.

## 2. The Model

As shown in the appendix, West et al.'s so called "universal equation" [19, 20] for organism (cell system) growth may be equivalently expressed as:

$$\frac{dm}{dt} = \frac{m}{C}\left[E\left(\frac{M}{m}\right)^{1/4} - E\right] \quad (1)$$

where $E$ is the natural death rate in the metabolism process and defined as the number of cells that died in a unit volume in unit time, $C$ the number density of the cells, $m$ is the instant mass ($m=VCm_c$, $V$ is the volume of the cell system and $m_c$ the mass of a single cell), $M$ is the asymptotically attained maximum mass within a system [19], though Guiot et al. [20] defined it as a final mass for a tumor. $E(M/m)^{1/4}$ represents the production rate (the average number of cells produced in a unit volume in unit time). Tumor growth under some conditions parallels that of a regular organism. Guiot et al. [20] applied West et al.'s equation to fit some data both in vitro and in vivo collected from literature for tumor growth.

When no radiation is applied, cells or organisms grow and die/shrink with a lower death rate when compared to the case where radiation is applied. Treatments cause the existence of an additional death rate. If the production rate is greater than the total death rate, a cell system or organism grows; otherwise, it shrinks or even disappears. When we apply eq. (1) to radiotherapy, the physical meaning is as follows: therapy causes an additional cell death rate $K$, which is defined as the number of cells "killed" in a unit volume in unit time. Thus, when treatment is applied, eq. (1) is amended to read:

$$\frac{dm}{dt} = \frac{m}{C}\left[E\left(\frac{M}{m}\right)^{1/4} - E - K\right] \quad (1')$$

This is a general expression for cell response to treatment. For radiation such as x-ray and α-particles, the dose in a unit time (dose rate) can be selected/controlled and set as a constant since the energy per particle and the distance of the target are fixed ($\dot{D}$ is set as a constant [21]). In a specific treatment period, the average dose rate is practically set as a constant. Therefore, the dose absorbed by an organism (or a cell system) is proportional to the amount of time it is exposed to radiation ($D=\dot{D}t$). We employ a general relation $t=\lambda D$, where $\lambda$ is the reciprocal of dose rate. Thus, eq. (1′) reads:

$$\frac{dm}{dD} = \frac{\lambda m}{C}\left[E\left(\frac{M}{m}\right)^{1/4} - E - K\right] \quad (2)$$

$K$ is a key parameter for determining cell survival response. A correct pattern of $K$ should reflect the experimental results. Considering the properties of α-particles, we divide it into three different cases:

a) When LET is greater than a specific value $L_0$ (LET>$L_0$) in a given cell system, the energy deposited is large enough to destroy all localized cells present equally (regardless whether the cells are hypoxic or oxygenated). The damage done to most cells is irreparable, or more accurately, the number of repairable damaged cells is much less than that of the irreparable ones. Therefore, the number of repairable cells may be ignored. The additional death rates are proportional to the quantity of cells present, which should be constant since the cell density is constant. In this case, the OER is equal to 1.0.

b) When LET is smaller than $L_0$ but greater than a given value $L_1$ ($L_1$<LET<$L_0$), the energy deposited is still large enough to cause serious damage immediately while interacting with cells and most damage is still irreparable. However, the energy deposited is not large enough to destroy all localized cells with the same effect. It is oxygen dependent. The damage under well-oxygenated condition is larger than that under hypoxic condition. The ratio of damage varies from one fold to nearly three folds. The exact value for a given cell system is LET dependent [1,3,5]. The number of cells killed is still proportional to the number of cells present, but the proportionality varies for cells under normoxic and hypoxic conditions. Therefore, the additional death rate is still constant (though the value is oxygen dependent) since the number density of cells is constant. In this case, the OER is greater than 1.0.

c) When LET is smaller than $L_1$ (LET<$L_1$), the energy deposited is not large enough to destroy all localized cells immediately. The mechanism of cell killing is the same as that for cells under low LET radiation. There is a time accumulating effect. We classify it as a linear-quadratic case.

In cases a) and b), the LET is large enough such that the localized nuclei of cells are damaged and the localized cells are destroyed immediately. There is no time accumulating effect. In this case, the additional death rate caused by α-particles is proportional to the cells present. For a specific kind of cell, the number density $C$ is fixed. Thus, the additional death rate is a constant ($K=v$). The solution of eq. (2) is:

$$E - (E+v)\left(\frac{m}{M}\right)^{1/4} = [E - (E+v)\left(\frac{M_0}{M}\right)^{1/4}]\exp(-\frac{E+v}{4C}\lambda D) \qquad (3)$$

The surviving fraction is

$$SF = \frac{m}{M_0} = \frac{M}{M_0(E+v)^4}\{E - [E - (E+v)(\frac{M_0}{M})^{1/4}]\exp(-\frac{E+v}{4C}\lambda D)\}^4 \qquad (3')$$

where $M_0$ is the mass of cells when radiation starts. In actuality, $M_0/M$ can be determined theoretically by comparing two identical cell systems. Radiation is applied to one but not the other. Usually, the additional death rate caused by alpha radiation is much greater than the production rate in a cell system, namely $(E+v)(m/M)^{1/4}>>E$ is satisfied. In this case, it gives an approximately linear cell survival curve and is expressed as

$$SF = \frac{m}{M_0} = \exp(-\frac{E+v}{C}\lambda D) \qquad (4)$$

Fig. 1 depicts log(*SF*) vs. *D* with various values of $E\lambda/(4C)$, $v\lambda/(4C)$ and $M_0$ based on eq. (3').

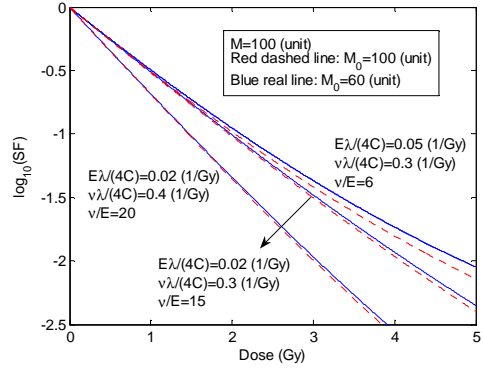

Fig. 1 General cell response to α-particle radiation with different parameters

For case c), the transferred energy is not large enough to damage some of the targeted nuclei immediately. The DNA damage of nuclei has a time accumulating effect. Since the energy deposited is not large enough, one attack does not cause serious damage to some cells. With time accumulating, the damage increases. Therefore, the damage of some cells is proportional to the time. When the damage reaches a certain amount, it causes cell death. Therefore, the additional death rate also has a time accumulating effect. Mathematically, the LQ equation (approximately expressed as $SF = N/N_0 = e^{-\alpha D - \beta D^2}$, where $N$ is the total number of cells and $N_0$ is the cell number at $D=0$) is equivalent to the equation $dN/N = -(\alpha + 2\beta D)dD$. The constant term $\alpha$ creates surviving fraction proportional factor $e^{-\alpha D}$, which is assumed to originate from single event killing. The linear term $2\beta D$ causes surviving fraction proportional factor $e^{-\beta D^2}$ and is assumed to be derived from double event killing [1, 16]. In fact, the linear term $2\beta D$ contains the time accumulating effect since $D = \dot{D}t$. Also, accumulation of DNA single strand break can cause DNA double helix break. Studies show that α-particles cause much more serious damage in the nuclei than low LET radiations, such as x-ray [1]. DNA double strand breaks are the primary critical lesion for cell killing [17, 18]. Considering these effects, we assume that some of the nuclei are damaged immediately, causing a constant additional death rate $\nu$, while others are damaged gradually under low LET radiation. With time or radiation dose accumulating, more and more DNA double strands are broken and nuclei are damaged correspondingly, which causes on average an additional death rate $\kappa D$. Therefore $K = \kappa D + \nu$, where $\kappa$ and $\nu$ are two constants for a specific radiation and target. Then eq. (2) becomes:

$$\frac{dm}{dD} = \frac{\lambda m}{C}\left[E\left(\frac{M}{m}\right)^{1/4} - E - \kappa D - \nu\right] \qquad (5)$$

The complete solution can be expressed by using the confluent hypergeometric function or given numerically. However, when condition $\kappa D + \nu + E >> E(M/m)^{1/4}$ is satisfied, eq. (5) is reduced to:

$$\frac{dm}{dD} = -\frac{\lambda}{C}m(\kappa D + \nu + E) \qquad (6)$$

The solution in such case is:

$$SF = \exp\{-\lambda[\tfrac{1}{2}\kappa D^2 + (\nu + E)D]/C\} \qquad (7)$$

Let $\alpha = (\nu+E)\lambda/C$, $\beta = \frac{1}{2}\kappa\lambda/C$, $\alpha/\beta = 2(\nu+E)/\kappa$, we have:

$$SF = \exp(-\alpha D - \beta D^2) \tag{7'}$$

This is the formula for the LQ model. For a specific target, the $\alpha$, $\beta$ and $C$ are fixed. Therefore, smaller $\lambda$ (corresponding to larger dose rate $\dot{D}$) causes larger $\kappa$ and $\nu$.

## 3. Application

Regarding cell response to α-particle radiation, the present model is compatible with all existing models. However, for C3H 10T1/2 cell response to α radiation given in reference [6], the experimental data cannot be fitted by any previous models, even considering the form for two component cell system. The energy of α-particles is 2.7Mev, their dose mean is LET 147 kev/μm, and frequency mean of LET is 144 kev/μm. When we apply eq. (3′) to the experimental data, it fits well. According to the description of the experiment, the abnormal experimental result cannot be explained due to the fact that unattached mitotic cells cannot be reached by the α-particles. Here we choose $M_0/M$=25% since the plating efficiency of control cultures in the experiment was 20-30%.

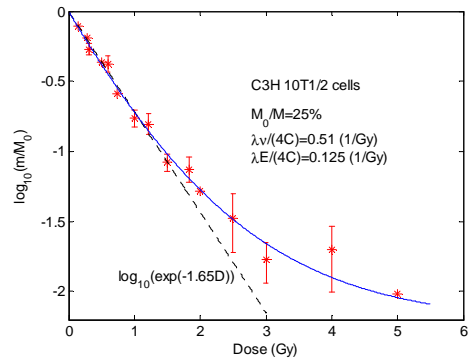

Fig. 2 C3H 10T1/2 cell response to α-particle radiation

Similarly, for A-549s's cell response to α-particle radiation in reference [7], the survival curve deviates away from the linear line. The last experimental point does not fit a linear response at all. However, eq. (3′) fits the experimental points well, which is shown in Fig. 3.

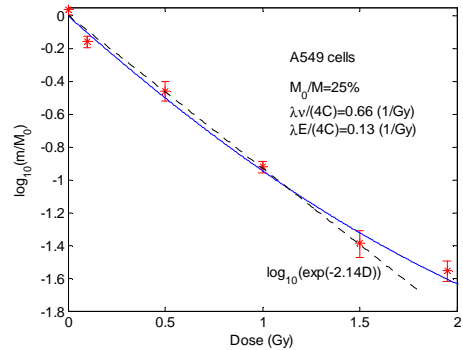

Fig. 3 A549s cell response to α-particle radiation

For H184B5 F5-1 M/10's cell response to α-particle radiation in reference [8], we can see that a linear line does not fit some experimental points well. Eq. (3′) fits the experimental points better, as show in Fig. 4.

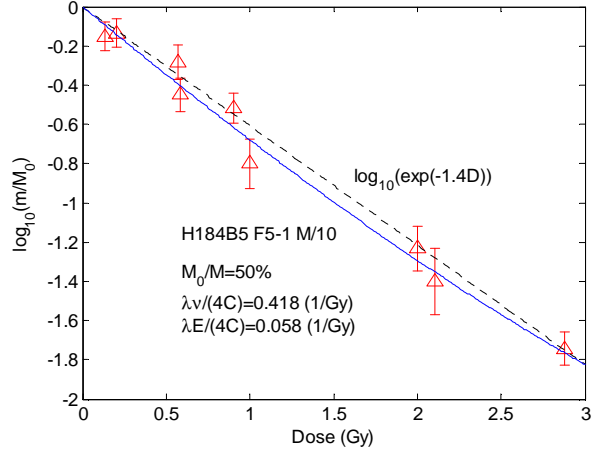

Fig. 4 H184B5 F5-1 M/10 cell response to α-particle radiation

For comparison, here we calculate the root-mean-square (RMS) differences based on the formula $RMS=\sqrt{\frac{1}{m}\sum_{i=1}^{m}(\log SF_i - \log SF_{fit})^2}$ [2]. We also calculate $\chi^2 = \sum_{i=1}^{m}\frac{(\log SF_i - \log SF_{fit})^2}{\sigma_i^2}$ if we know the variance of each experimental point, where $\sigma_i$ represents the variance of $\log SF_i$. We do so since the increments of $\log SF$ are equally spaced and the differences of $\log SF_i - \log SF_{fit}$ are "more easily correlated to the goodness of fit that is visually apparent in the usual semi-log plot of $SF$ vs. $D$ [2]". The results are listed in table 2. It shows that the present model fits the curves better.

**Table 2 The goodness of linear fit and the fit of the present model**

| Cells | The present model (free fit) | | The linear fit | |
|---|---|---|---|---|
| | RMS | $\chi^2$ | RMS | $\chi^2$ |
| A549s | 0.0464 | 5.69 | 0.110 | 22.28 |
| H184B5 | 0.0730 | 5.416 | 0.0888 | 5.717 |

Besides using α-particles, Barendsen et al. [3] also used deuterons and neutrons to irradiate T-1g cells in culture under oxygenated condition (with air) and hypoxic condition (with nitrogen). For deuterons, when the LET is 20kev/μm, the survival curves are almost straight lines. However, when the LET is 5.6kev/μm, they are not straight anymore. Here we apply the LQ model (eq. 7′) to fit the experimental data. The result is shown in Fig. 5. Similarly, for the survival curves under 25 Mev α-particles with LET 26±2kev/μm, they cannot be fitted linearly, especially for the survival curve under hypoxic condition. We use the LQ model to fit the survival curves. The result is shown in Fig. 6. Obviously, it shows that the survival curves' response to heavy particles such as α-particles and deuterons also have the characteristics of x-ray when LET is low.

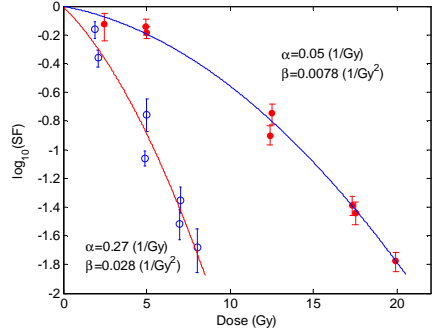

Fig. 5 T-1g cell response to deuterons (14.9Mev) with air (open circles) and nitrogen (closed circles). LET: 5.6±0.3kev/μm; OER: 2.6±0.3

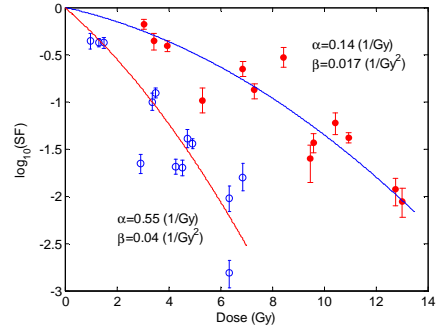

Fig. 6 T-1g cell response to α-particles (25Mev) with air (open circles) and nitrogen (closed circles). LET: 26±2kev/μm; OER: 2.4±0.3

## 4. Discussion

As we mentioned above, the additional death rate $K$ should be the difference between the number of damaged and repaired (self regeneration) cells in a unit volume in a unit time. There are three cases: 1) LET is large. In this case, most damaged cells are irreparable and there is no difference between hypoxic and oxygenated. 2) LET is medium. Though the damage is mostly irreparable, the number of damaged cells is oxygen and LET dependent. Since the LET is high enough to cause a serious damage to the cells present immediately in these two cases, there is no time accumulating effect. Generally, the amount of cells damaged in a unit volume in unit time is the sum of the number of irreparable and repairable cells in a unit volume in unit time. It is expressed as $N_d=N_i+N_r$. The number of repaired cells should be proportional to the number of repairable ones. Therefore, the number of cells repaired in a unit volume in unit time is $n_r=cN_r$, where $c$ is the proportionality constant. The additional death rate should be the difference between the number of cells in a unit volume in unit time that are damaged and that of repaired cells. This is expressed as $K=N_d-n_r=N_i+(1-c)N_r$, which is not time or dose related. The larger the LET, the larger the number of irreparably damaged cells, $N_i$. If LET does not change, $N_i$ is larger under aerobic condition than under hypoxic condition. When LET>$L_1$, $N_i \gg N_r$. Most damaged cells are irreparable and the number of cells that can be repaired may be ignored. When LET>$L_0$, the energy deposited is large enough to kill all localized cells with the same effect. There is no difference in cell damage between hypoxic and aerated condition. 3) LET is low. In this case, some cells are damaged by radiation gradually. The number of damaged cells has a time accumulating effect: $n_d=\nu_d+\delta_d t$ ($\nu_d$ includes irreparable and repairable damaged cells). Considering $t=\lambda D$, we rewrite: $n_d=N_i+N_r+\kappa_d D$, which represents number of damaged cells in unit volume in unit time. The number of repaired cells should be proportional to the number of repairable damaged ones: $n_r = c_1 N_r + c_2 \kappa_d D$, $c_1$ and $c_2$ are two proportionality constants. Since these two terms are derived from distinct killing mechanism, $c_1$ and $c_2$ are not the same ($c_1 \neq c_2$). Figs. 5 and 6 also show that the values of $\alpha$ and $\beta$ are not proportional when cells are under hypoxic and oxygenated conditions. Therefore, the additional death rate is $K = n_d - n_r = \kappa D + \nu$. According to the discussion above, $\kappa$ and $\nu$ (correspondingly $\alpha$ and $\beta$) change when the oxygenated condition varies. How to determine the change pattern needs further study.

If the cell system is composed of two different kinds of cells (one resistant, the other regular), the relationship between the surviving fraction and radiation dose is complicated. It is easy to show that the fractions of different cells in a system before therapy ($K=0$) are constant if various cells have the same $E/C$ and their asymptotic masses share the same fractions. These two conditions ($E_i/C_i$ is constant and $M_i = x_i M$) ensure that the cells grow harmoniously before treatment. The different responses to therapy enable us to detect the existence of various cell subpopulations. Assuming that the resistant cells have $x$ fraction (regular cells $1-x$ fraction). For a system composed of identical cells, the mass is proportional to the volume or the number of cells ($m = V C m_c = N m_c$). The surviving fraction of the entire system is $SF = m/M_0 = (m_{res} + m_{reg})/M_0$. If the LET is large enough to kill all localized cells equally, the additional death rates are proportional to the amount of cells present and have a relationship of $\nu_{res}/C_{res} = \nu_{reg}/C_{reg} = \theta$ for different kinds of cells. For cells in the same tissue or from the same source which grow harmoniously before therapy, condition $E_{res}/C_{res} = E_{reg}/C_{reg} = \beta^\circ$ should be satisfied. Also, $M_{res} = xM$ and $M_{0res} = xM_0$. Thus we have,

$$SF = \frac{m_{res} + m_{reg}}{M_0} = \frac{M}{M_0 (\theta + \beta^\circ)^4} \{\beta^\circ - [\beta^\circ - (\theta + \beta^\circ)(\frac{M_0}{M})^{1/4}]\exp[-(\theta + \beta^\circ)\lambda D/4]\}^4 \quad (8)$$

In this case, the consequence is the same as eq. (3′) for single cell component system. When the total death rates are much greater than the production rates for both cell populations, we have:

$$SF = \exp[-(\theta + \beta^\circ)\lambda D] \quad (8')$$

Eq. (8′) shows a linear relation of $\log(SF) \sim D$. For cells that do not grow harmoniously before therapy, we have to use eq. (3) for calculating the masses separately and then obtain the total surviving fraction of cells $SF = (m_{res} + m_{reg})/M_0$. Since $M$, $M_0$ and $\beta^\circ$ are not related for different cell populations, the situation is complicated. If LET is not large enough, $\nu_{res}/C_{res} \neq \nu_{reg}/C_{reg}$ since the number of reparable cells for one or even both subpopulations cannot be ignored. In this case, eq. (8) may not be satisfied. If LET is low, the total surviving fraction may be determined by using the two-component LQ model [2]. In this case, how should the values of $\alpha$ and $\beta$ for different types of cells be constrained for the best fit? Also, we find that some survival curves of cell response to x-ray *in vivo* can only be fitted by using eq. (5). We will not discuss this in detail here. Further studying is needed.

## 5. Conclusion

Therapies cause an additional death rate (relative to the natural death rate). The present model connects the consequence of radiation with the growth process of a biosystem. This model explains the response of cell population under both high and low LET radiations consistently. Traditionally, high and low LET radiations are classified by the types of ions or particles of radiation. Usually, x-ray, γ-ray and β-particles are categorized as low LET radiations, whereas α-particles, protons, neutrons etc. (heavy particles) are taken as high LET radiations. Based on this model, we now suggest that the categories of high and low radiations should be classified according to the exact values of LET. When LET is high enough, radiation causes a linear survival response; while LET is low, the cell response follows the linear-quadratic pattern. The present model explains that the oxygen-enhancement ratio (OER) decreases while LET increases consistently. It also shows that the LQ model is an approximation of the present model under a specific condition: the total death rate, which includes additional death rate caused by radiation and natural death rate, is much greater than the production rate.

## Appendix: The equivalent expression of West et al.'s universal equation for organism growth

West et al. [19] proposed a model for organism growth, which is expressed as:

$$\frac{dm}{dt} = am^{3/4}\left[1-\left(\frac{m}{M}\right)^{1/4}\right] = aM^{-1/4}m\left[\left(\frac{M}{m}\right)^{1/4}-1\right] \quad (1)$$

Tumor growth under some conditions parallels that of an organism. Guiot et al. [20] applied this equation to fit some data both in vitro and in vivo collected from literature for tumor growth. In fact, if we define the production rate $\bar{S}$ as the average number of cells produced in a unit volume in unit time, the death rate $E$ as the number of cells that died in a unit volume in unit time, then the growth rate of a system which is composed of identical cells is:

$$\frac{dm}{dt} = m_c V(\bar{S}-E) = \frac{m}{C}(\bar{S}-E) \quad (2)$$

where $m_c$ is the mass of a single cell, $m$ is the instant mass, $V$ the volume and $C$ the number density of the cells ($m=VCm_c$). Since "the cell death rate is proportional to the number of cells present" [19], it is reasonable to assume that the death rate $E$ of cells is a constant in this case. West et al. indicated that "rates of cellular metabolism and heartbeat [scale] as $M^{-1/4}$ and whole-organism metabolism rate as $M^{3/4}$." (G. B. West et al., Science 284 p1677). Considering the general ¼ power law in biological allometry, the production **rate** should be the reciprocal of the ¼ power of the mass, namely, $\bar{S} \propto 1/m^{1/4}$ or $\bar{S} = \mu m^{-1/4}$, where $\mu$ is a proportionality factor. Also, considering the condition: $m=M$ at $dm/dt=0$, we have $\mu = EM^{1/4}$. Therefore, $\bar{S} = E\left(\frac{M}{m}\right)^{1/4}$, where $M$ is the asymptotically attained maximum mass among species within a system [19], though Guiot et al. [20] defined it as a final mass for a tumor. By substituting it in eq. (2), we have:

$$\frac{dm}{dt} = \frac{E}{C}m\left[\left(\frac{M}{m}\right)^{1/4}-1\right] \quad (3)$$

In reference [19], West et al. discussed the death rate of cells and obtained that death rate in total volume is $aM^{-1/4}N_c$ ($N_c$ is the total cell number). It should be equal to $EV$ in this case: $EV = aM^{-1/4}N_c$. Since $C=N_c/V$, $\frac{E}{C} = aM^{-1/4}$. Therefore, eqs. (1) and (3) are the same.